\documentclass[fleqn,setspace,amsmath,amssymb]{article}
\usepackage[dvips]{graphicx}
\usepackage{setspace}
\usepackage{amsmath}
\usepackage{amssymb}
 
\newcommand{\ra}{\rangle}

\newcommand{\la}{\langle}
\newcommand{\be}{\begin{equation}}
\newcommand{\ee}{\end{equation}}
\newcommand{\bea}{\begin{eqnarray}}
\newcommand{\eea}{\end{eqnarray}}

\begin{document}
\begin{titlepage}

\begin{flushright}
\today
\end{flushright}

\vspace{1in}

\begin{center}

{\bf Mechanism for coherence in the resonant system of ion-solvated water molecules and radiation}

\vspace{1in}

\normalsize

{Eiji Konishi\footnote{E-mail address: konishi.eiji.27c@kyoto-u.jp}}

\normalsize
\vspace{.5in}

{\it Graduate School of Human and Environmental Studies, Kyoto University\\
 Kyoto 606-8501, Japan}
\end{center}

\vspace{1in}

\baselineskip=24pt
\begin{abstract}
This paper presents a comprehensive exposition of a spontaneous laser model for a resonant semi-classical system of radiation and ion cluster-solvated rotating water molecules, which have subtly variable moments of inertia.
In this system, ions in the cluster carry the same electric charge and move with very low, non-relativistic velocities in a direction parallel to an applied unidirectional static electric field.
The role of the static electric field is to induce electrostatic mixing of the rotational states of the water molecules.
We assume that the dimensions of the ion cluster are much shorter than the wavelength of the radiation in the resonant interaction.
In this model, we describe rotating water molecules quantum mechanically by using a two-level approximation, and we show that the equations of motion of the system are the same as those of a conventional free electron laser system.
This result and the existence of permanent electric polarization of the water molecules by electrostatic mixing lead to a mechanism for radiation coherence induced by collective instability in the wave-particle interaction.
As an illustrative example, we apply this mechanism to action potential propagation in myelinated neuronal axons of the human brain.
\end{abstract}

\vspace{.7in}
 
\end{titlepage}

\tableofcontents

\section{Introduction}

In quantum physics, {\it coherence} is a mainstay concept, sharing its importance with {\it entanglement}.
After the appearance of the decoherence program\cite{Decoherence,Decoherence2,Decoherence22,Decoherence3,Decoherence23}, it was noticed that a quantum-mechanical pure state with coherence is a non-trivial state in macroscopic open quantum systems.
It is, therefore, worthwhile to propose a general spontaneous mechanism for the generation of quantum coherence.

In low-energy physics, specifically, quantum and classical electrodynamics, the spontaneous mechanisms behind the coherence of radiation are mainly classified into two types.
In the first type, coherence is quantum mechanically based and due to the cooperative quantum state of the radiators.
In the second type, coherence is due to the collective instability, in a many-body system with long-range interactions, which magnifies the radiation intensity and bunches the radiators' phases.
In spontaneous laser models, Dicke superradiation belongs to the first type\cite{Dicke,Super} and the free electron laser (FEL) belongs to the second type\cite{FEL,Bonifacio,FELPR,FEL1,FEL3}.

In this paper, we present a comprehensive exposition of a new spontaneous laser model that was proposed previously in shorter reports by the author\cite{Konishi,Konishi2} and incorporates both types of mechanism for coherence of radiation from the aspect of wave-particle interactions\cite{Book}.

Specifically, we consider a model for a cluster of ions, solvated by water molecules, with the same electric charge, very low, non-relativistic velocities, and a uniform direction of motion parallel to an applied unidirectional static electric field.
The role of this electric field is to induce electrostatic mixing of the rotational states of the water molecules.\cite{GPV}
We assume that the dimensions of the ion cluster are much shorter than the wavelength of the radiation in the resonant interaction.

In this model, we regard a water molecule as a quantum mechanical rigid body with a moment of inertia, rotating about its electric dipole moment axis, that is subtly {\it variable} because of a variable internal structure, that is, the vibrational modes of the water molecule.
Then, the electric dipole moment of each ion-solvated water molecule can be described by an $XY$ energy spin under a two-level approximation of the subtly varying rotational spectrum of the water molecules.
This $XY$ energy spin, that is, an ion-solvated water molecule interacts resonantly with the transverse electromagnetic field, of a {\it specific mode} (i.e., a specific pair of a wavevector and a polarization), that is radiated from and absorbed by the water molecules in certain rotational states.

Note that, on the particle side of the wave-particle (i.e., radiation field-particle) interactions in this model, there are two kinds of elements, namely, water molecules and ions, where each ion is coupled electrostatically to (i.e., quantum mechanically entangled with) $30\sim 35$ water molecules around it\cite{SB}.
We use two main ideas to describe the water molecule states:
\begin{enumerate}
\item Because each moving ion can be described by a {\it pure state}, the ion-solvated water molecules can also be described by {\it pure states}.
\item By contrast, the bulk water molecules cannot be described by pure states and are assumed to be in a canonical distribution.
\end{enumerate}

Based on this setup and these two ideas, the specific goal of our investigation is to show that the equations of motion of the $XY$-phases of ion-solvated water molecules, over all ions, and the transverse electromagnetic field are the same as the equations of motion of the conventional FEL model that exhibits a mechanism for radiation coherence\cite{Nature}.
Then, in the presence of permanent electric polarization of the water molecules due to electrostatic mixing\cite{GPV}, the quantum laser mechanism works.

This paper is organized as follows.
In Section 2, we describe the quantum state of a single water molecule under a two-level approximation for its rotational spectrum.
In Section 3, we make four assumptions and explain their consequences.
In Section 4, we derive the Hamiltonian for the resonant interaction between the ion-solvated water molecules and the radiation field of a specific mode in a semi-classical treatment.
In Section 5, we derive the equations of motion of the system.
In Section 6, we show that these equations of motion are the same as those of a conventional FEL system, explain the FEL-like mechanism, that is, the collective instability schematically, and derive the formulae for the gain intensity of the radiation and the gain time (that is, the dynamical time) arising from this mechanism in our system.
In Section 7, we apply our mechanism to action potential propagation in myelinated neuronal axons of the human brain as an illustrative example.
Finally, we summarize the overall results in Section 8.

\section{Description of a Single Water Molecule}

\subsection{Two-level approximation and state space}

In this paper, {\it water} is regarded as an ensemble of {\it quantum mechanical rigid rotators}, that is, water molecules having a subtly {\it variable} moment of inertia, whose average value is $I^{(w)}_{\rm ave}=2m_pd_g^2$.
Here, $m_p$ is the proton mass and $d_g\approx 0.82$ [\AA]\cite{Franks}.

In addition to this picture of water, our description is based on the assertion that, to a good approximation in our system, the radiation field exchanges energy with the water molecules only through excitation and de-excitation between the two lowest levels of the internal rotation of the hydrogen atoms about the electric dipole axis of each water molecule.

This assertion is supported by two facts.
First, the population ratios $R_l$ between the water molecules in each of the rotational states $\{|l,m\ra|-l\le m\le l\}$, where $l$ and $m$ refer to the azimuthal quantum number and the magnetic quantum number, respectively, and those in the rotational state $|0,0\ra$ are
\begin{equation}
R_0=1\;,\ R_1=0.89\;,\ R_2=0.70\;,\ R_3=0.49\;,\ R_4=0.30\;,\ R_5=0.17\;,\ \ldots
\end{equation}
in thermal equilibrium at room temperature $300$ [K].\cite{GPV}
Second, the probability of radiation-induced transition between the two lowest rotational energy states is high relative to that between two rotational energy states of any other allowed pair.
These two facts justify ignoring the radiation-induced transition processes between two rotational energy states of any other allowed pair.

The energy difference between the lowest two levels for the {\it average} moment of inertia $I^{(w)}_{\rm ave}$ is\cite{GPV,Franks}
\begin{equation}
{\cal E}=\frac{\hbar^2}{I^{(w)}_{\rm ave}}\;,\ \ \frac{{\cal E}}{\hbar c}\approx 160\ [{\rm cm}^{-1}]\;.
\end{equation}

In this two-level approximation of the rotational spectrum of the water molecules, the Hilbert space ${\cal V}$ of the internal rotational states of a water molecule is four-dimensional.
It is spanned by the state $|l,m\ra$ with $l=0$ and $m=0$ and the three states $|l,m\ra$ with $l=1$ and $m=1,0,-1$:
\begin{equation}
{\cal V}=\la |0,0\ra,|1,1\ra,|1,0\ra,|1,-1\ra\ra_{{\mathbf C}}\;.
\end{equation}
In this Hilbert space ${\cal V}$, the off-diagonal electric dipole moment operator vector $\vec{\widehat{d}}$\footnote{In this paper, a hat indicates that a variable is a quantum mechanical operator.} does not interchange any two different states $|1,m_1\ra$ and $|1,m_2\ra$, and the component operators $\widehat{d}^1$, $\widehat{d}^2$ and $\widehat{d}^3$ interchange the {\it ground} state $|g\ra\equiv|0,0\ra$ with the states
\begin{eqnarray}
|d^1\ra&=&\frac{-(|1,1\ra-|1,-1\ra)}{\sqrt{2}}\;,\\
|d^2\ra&=&\frac{i(|1,1\ra+|1,-1\ra)}{\sqrt{2}}\;,\\
|d^3\ra&=&|1,0\ra\;,
\end{eqnarray}
respectively.

We take advantage of the rotational symmetry of the problem.
Specifically, when we describe the ponderomotive potential of each ion system created by the radiation field after solving the internal dynamics of water molecules in each ion system, we truncate the electric dipole moment operator vector $\vec{\widehat{d}}$ of each water molecule by the projection of the vector $(|d^1\ra,|d^2\ra,|d^3\ra)$ onto the one-dimensional space generated by the {\it excited state} $|e\ra\equiv |1,1\ra$:
\begin{equation}
(|d^1\ra,|d^2\ra,|d^3\ra)\to \biggl(\frac{|d^1\ra+i|d^2\ra}{2},\frac{|d^2\ra-i|d^1\ra}{2},0\biggr)\;.
\end{equation}
Furthermore, we treat only the specific electromagnetic mode coupled to the transitions between $|g\ra$ and $|e\ra$.

For each water molecule, we denote its electric dipole direction vector by $\vec{e}_3$.
To truncate the four-dimensional Hilbert space of that water molecule as
\begin{equation}
{\cal V}_{tr}=\la |g\ra,|e\ra \ra_{{\mathbf C}}\;,
\end{equation}
we choose its quantization axis to lie along $\vec{e}_3$.

\subsection{Energy spin variables and the free Hamiltonian}

To describe the truncated electric dipole moment operator, we introduce {\it energy} spin variables for each water molecule:
\begin{eqnarray}
\widehat{s}^1&=&\frac{1}{2}[|e\ra\la g|+|g\ra\la e|]\;,\\
\widehat{s}^2&=&\frac{1}{2i}[|e\ra\la g|-|g\ra\la e|]\;,\\
\widehat{s}^3&=&\frac{1}{2}[|e\ra\la e|-|g\ra\la g|]\;.\label{eq:Spin}
\end{eqnarray}
Here, as introduced above, $|g\ra$ and $|e\ra$ are, respectively, the ground state $|0,0\ra$ and a first excitation energy state $|1,1\ra$ of the water molecule in the two-level approximation.
The superscripts of the energy spins represent fictitious dimensions\cite{Dicke,PRSR}.
These three energy spin operators obey an $su(2)$ algebra $[\widehat{s}^i,\widehat{s}^j]={\rm i}\epsilon_{ijk}\widehat{s}^k$.

For each water molecule, the free Hamiltonian defined in the Hilbert space ${\cal V}$ in the two-level approximation of the rotational spectrum is
\begin{equation}
\widehat{H}_{\rm free}=\frac{1}{2}{\cal E}\widehat{I}_{3,1}\;,\label{eq:Hfree}
\end{equation}
where
\begin{equation}
\widehat{I}_{3,1}\equiv [|1,1\ra\la 1,1|+|1,0\ra\la 1,0|+|1,-1\ra\la 1,-1|-|0,0\ra\la 0,0|]\;.
\end{equation}
Whereas the original electric dipole moment operator ${\widehat{d}}$ transforms as a {\it vector} with respect to spatial rotations, the operator $\widehat{I}_{3,1}$ is a {\it scalar} operator.
Since $\widehat{I}_{3,1}$ is a diagonal operator, it cannot be a component of a spatial vector.
So, we truncate the electric dipole moment operator but do not truncate the free Hamiltonian.
Note that $[\widehat{I}_{3,1}/2,\widehat{s}^a]=[\widehat{s}^3,\widehat{s}^a]$ holds for $a=1,2,3$.

\subsection{Truncated electric dipole moment operator}

We have chosen the third axis of the electric dipole moment operator vector $\vec{\widehat{d}}$ for a water molecule as the quantization axis for its rotation.

One of the off-diagonal matrix elements $\la e|\vec{\widehat{d}}|g\ra=\overline{\la g|\vec{\widehat{d}}|e\ra}$ of the (truncated) electric dipole moment operator vector $\vec{\widehat{d}}$ of this water molecule is given by
\begin{eqnarray}
&&d_0\int_{-1}^1d(\cos \theta)\int_0^{2\pi}d\varphi (\bar{Y}_{1,1}(\theta,\varphi)(\vec{e}_1\sin\theta \cos \varphi+\vec{e}_2\sin \theta \sin \varphi+\vec{e}_3\cos \theta)Y_{0,0}(\theta,\varphi))\nonumber\\
&&={d_0}\sqrt{\frac{1}{6}}(-\vec{e}_1+i\vec{e}_2)\;.\label{eq:d0}
\end{eqnarray}
Here, we choose the zenith angle as $\theta$, and $d_0=2ed_e$ with $d_e\approx 0.2$ [\AA]\cite{Franks}.
The electric dipole moment operator vector can then be represented as an off-diagonal matrix vector in the two-dimensional energy state space ${\cal V}_{tr}$:\cite{Dover}
\begin{equation}
\vec{\widehat{d}}=\frac{\widetilde{d}_0}{2}(-\vec{e}_1(2\widehat{s}^1)-\vec{e}_2(2\widehat{s}^2))\;.\label{eq:dex}
\end{equation}
Here, we define
\begin{equation}
\widetilde{d}_0=d_0\sqrt{\frac{2}{3}} \approx 0.82\cdot d_0\;.
\end{equation}
Expression (\ref{eq:dex}) for the electric dipole moment operator will be used in Section 4.2.

\section{Setup}
\subsection{Assumptions}

For the system treated herein, we make the following four assumptions.

\begin{enumerate}

\item[A1] We apply a static electric field $\vec{E}_0$ in the $z$-direction, and then ions move along the $z$-axis with velocity $v\ll c$.\footnote{In this paper, we ignore quantities of the order of $(v/c)^n$ ($n\ge 1$).}
This static electric field $E_{0,z}$ induces the permanent electric polarization, $P_z$, of the water molecules.\cite{GPV}
Here, the {\it permanent electric polarization} is the time-independent part of the induced electric polarization, where the rest part vanishes after the temporal arithmetic averaging.\cite{GPV}

\item[A2] The dimensions of the ion cluster are much shorter than the wavelength of the radiation, the latter being denoted by $l_c$, so that the spatial part of the phase of the radiation field is approximately the same at the spatial sites of different ions.
Here, this wavelength $l_c$ is the inverse of the wavenumber of the radiation in the resonant interaction ${\cal E}/(hc)$.
Because ${\cal E}/(\hbar c)\approx 160$ [cm$^{-1}$], we have $l_c\approx 400$ [$\mu$m].\cite{Franks}

\item[A3] The ion-solvated water molecules can be described by a quantum mechanical {\it pure state}.

\item[A4] The bulk water molecules are in a quantum mechanical {\it mixed state}, specifically, the canonical distribution (i.e., the thermal equilibrium state).

\end{enumerate}

Note that long-wavelength radiation interacts weakly with water molecules.
However, in our mechanism, the water molecules interact {\it coherently} with the selected mode of the radiation, similar to the interactions between relativistic electrons and radiation in the FEL model.

\subsection{Water molecule states}

First, we clarify the main ideas in the description of the water molecule states by making the following three points.
These three statements are the grounds for assumptions A3 and A4.

\begin{enumerate}

\item[S1] In the solution, each ion moves and can be described by a pure state.

\item[S2] Then, the rotational states of the water molecules coupled electrostatically to each ion can also be described by pure states.

\item[S3] By contrast, the rotational states of water molecules in the bulk system that are decoupled from ions cannot be described by pure states; instead, they are described by a canonical distribution.

\end{enumerate}

Using S1 and S2, we determine the forms of the ion-solvated water molecule states.

\subsubsection{Ion-solvated water molecules}

From the quasichemical theory analysis, it is found that the electrostatic effect of an ion (Li$^+$, Na$^+$, K$^+$, Cs$^+$, F$^-$, Cl$^-$, Br$^-$, I$^-$) on solvating water molecules reaches a distance of $6.15$\AA, and within this shell, there are $30\sim 35$ water molecules.\cite{SB}
This result means that $n=30\sim 35$ water molecules are electrostatically coupled to (i.e., quantum mechanically entangled with) each ion.

We set $\Delta n=n_--n_+$, where $n_+$ and $n_-$ refer, respectively, to the numbers of excited and ground state water molecules coupled to one ion, arithmetically averaged over all ions.
Here, we assume thermal equilibrium (i.e., the canonical distribution) and obtain the thermally averaged value
\begin{eqnarray}
\overline{\Delta n}\approx\frac{n}{2}\tanh\biggl(\frac{{\cal E}}{2k_BT}\biggr)\;,\label{eq:azi}
\end{eqnarray}
where $n/2=n \dim_{\mathbf C}{\cal V}_{tr}/\dim_{\mathbf C}{\cal V}$, ${\cal E}/2k_BT\approx 0.06$ at room temperature ($T=300$ [K]), and $\overline{\Delta n}\approx 0.9$ for $n=30$.

The time evolution law of the water molecules in the system, which defines the coupling between water molecules and the radiation field, is symmetrized with respect to permutations of water molecules.
This is because a radiation wavelength of the order of $l_c$ is much longer than the dimensions of the system of one ion and its solvent water molecules, that is, of the order of $1$ [nm]\cite{SB}; we call this fact $\star$.\cite{PRSR}

With respect to the $j$-th water molecule ($j\in \{1,2,\ldots,n\}$) coupled to (i.e., quantum mechanically entangled with) the $I$-th ion, the most general forms of the internal parts of the wave functions, $\psi^{(g)}_j$ and $\psi^{(e)}_j$, of the truncated, therefore not normalized, {\it superradiant states} excited and de-excited, respectively, from the rotationally symmetric disorganized thermal equilibrium into a set of rotationally symmetric collective energy spin states are
\begin{eqnarray}
|\psi^{(g)}_j\ra&=&\frac{1}{\sqrt{2}}\biggl\{\frac{1}{\sqrt{3}}|e\ra_je^{i\vartheta_1}-|g\ra_je^{i\vartheta_2}\biggr\}\;,\label{eq:psig} \\
|\psi^{(e)}_j\ra&=&\frac{1}{\sqrt{2}}\biggl\{|g\ra_je^{-i\vartheta_1}+\frac{1}{\sqrt{3}}|e\ra_je^{-i\vartheta_2}\biggr\}\label{eq:psie}
\end{eqnarray}
with the lowest cooperation number in Dicke's definition\cite{Dicke} and the specific (i.e., unaveraged) number $(3n_+-n_-)/2$ having been reduced to zero by a classical radiation pulse at the $I$-th ion.
Here, because of fact $\star$, we can set
\begin{equation}
\vartheta_2-\vartheta_1=\theta_I+\delta\label{eq:origin}
\end{equation}
with $\delta=\omega_ct$ for resonance angular frequency $\omega_c=2\pi c/l_c$.

In our description of each ion system, if the moment of inertia of the water molecules can vary dynamically from its average value $I^{(w)}_{\rm ave}$ (see Eq.(\ref{eq:Iwchange0})), then $\theta_I$ is dynamical and can have $I$-dependence.
Otherwise, $\theta_I$ is auxiliary, has no non-trivial $I$-dependence, and is compatible with monochromatic plane-wave radiation solutions only (see Section 5.1).

\subsubsection{Bulk water molecules}

The bulk water molecules that screen the electric charges of ions do not form part of our laser mechanism (see Section 5.2); the bulk water molecules are in a {\it mixed state}, assumed to be the truncated canonical distribution
\begin{equation}
\widehat{\varrho}^{(w)}_{\rm bulk}=\bigotimes_{{\rm WM}\in {\rm bulk}}(f_+(\beta)|e\ra\la e|+f_-(\beta)|g\ra\la g|)\;.\label{eq:candis}
\end{equation}
Here, $\beta=1/k_BT$ is the inverse temperature, and $f_+(\beta)$ and $f_-(\beta)$ are the Boltzmann weights of the single-water molecule states $|e\ra$, a first excitation energy state, and $|g\ra$, the ground state, respectively.

Consequently, this mixed state has {\it no} quantum coherence with respect to the energy eigenstates $|g\ra$ and $|e\ra$, and the relative phases between the energy eigenstates $|g\ra$ and $|e\ra$ in this mixed state of the bulk water molecules are uniformly randomly distributed.

\subsection{Permanent electric polarization}

In Ref.\cite{GPV}, a resonant system of rotating water molecules, whose electric dipoles interact with the radiation field, was studied using the path-integral approach of quantum field theory.
It was supposed that a static electric field, $\vec{E}_0$, oriented in the $z$-direction is applied.
Then, it was shown that, in the limit cycle of the system, a permanent electric polarization $P_z$ in the $z$-direction of water molecules in the stationary configuration, where the quantization axis of the rotational states $|l,m\ra$ is chosen as the $z$-axis, is induced by the electrostatic mixing of two rotational states $|0,0\ra$ and $|1,0\ra$ of water molecules in two new rotational eigenstates
\begin{eqnarray}
|\widetilde{0}\ra&=&\cos \alpha|0,0\ra+\sin\alpha |1,0\ra\;,\label{eq:mix1}\\
|\widetilde{1}\ra&=&-\sin\alpha|0,0\ra+\cos\alpha |1,0\ra\;,\label{eq:mix2}
\end{eqnarray}
where $\alpha$ is negligibly small but non-zero.
This electrostatic mixing is the perturbation of the rotational eigenstates $|0,0\ra$ and $|1,0\ra$ by the electrostatic potential
\begin{eqnarray}
V_d&=&-d_0E_{0,z}\cos \theta \\
 &=&-d_0E_{0,z}\sqrt{\frac{4\pi}{3}}Y_{1,0}(\theta)
\end{eqnarray}
induced by $\vec{E}_0$.

The intuitive explanation for the non-zero $P_z$ is as follows.
The {\it electric polarization} of water molecules in the $z$-direction is quantum mechanically the expectation value of the $z$-component of the electric dipole moment direction vector $\vec{e}_3$ taken with respect to the rotational state of the water molecules.
Here, the electric dipole moment direction vector $\vec{e}_3$ has non-zero matrix elements only between two rotational states with different azimuthal quantum numbers $l=0$ and $l=1$, that is, the off-diagonal matrix elements in the energy eigenbasis.
So, if there is no electrostatic mixing, then a temporal oscillation factor accompanies the off-diagonal matrix elements, and the off-diagonal matrix elements vanish under temporal averaging.
However, the electrostatic mixing in Eqs.(\ref{eq:mix1}) and (\ref{eq:mix2}) produces off-diagonal matrix elements without temporal oscillation and gives a non-zero permanent electric polarization $P_z$.

Explicitly, since we choose the $z$-axis as the quantization axis of $|0,0\ra$ and $|1,0\ra$, we obtain
\begin{eqnarray}
P_z&=&\frac{w_-+3w_+}{2}\la \widetilde{0}|\vec{e}_3\cdot \vec{e}_z|\widetilde{0}\ra+\frac{w_-+3w_+}{6}\la \widetilde{1}|\vec{e}_3\cdot \vec{e}_z|\widetilde{1}\ra \label{eq:PzEx}\\
&=&\frac{1}{3}\la 1,0|\vec{e}_3\cdot\vec{e}_z|0,0\ra \sin 2\alpha \\
&=&\frac{1}{3\sqrt{3}}\sin 2\alpha\;,\label{eq:GPVResult}\\
P_x&=&0\;,\\
P_y&=&0\;,
\end{eqnarray}
where $w_-$ and $w_+$ are the {\it average} population proportions of water molecules in the energy eigenstates $|\widetilde{0}\ra$ and $|\widetilde{1}\ra$, respectively.
Note that $w_-+3w_+=1$ holds and $(P_x,P_y,P_z)$ is a vector.

In the limit cycle of the system, $\Delta w=w_--w_+$ is obtained as\cite{GPV}
\begin{equation}
\Delta w \approx 0.62\;,\label{eq:Delw}
\end{equation}
where the system is initially in thermal equilibrium.
This result changes $\overline{\Delta n}$ from Eq.(\ref{eq:azi}) to
\begin{eqnarray}
\overline{\Delta n}&=&n \Delta w \\
&\approx&18.7\;.
\end{eqnarray}

Applying the formula (\ref{eq:Delw}) and the formula for $\alpha$ in Ref.\cite{GPV} to Eq.(\ref{eq:GPVResult}), we obtain the dependence of $P_z$ on $E_{0,z}$ for a realistic value of $E_{0,z}$:
\begin{eqnarray}
P_z&\approx&c_P\cdot E_{0,z}\;,\label{eq:Pzf1}\\
c_P&\approx&4.9\cdot 10^{-9}\ [{\rm V}^{-1}\cdot{\rm m}]\;.\label{eq:Pzf2}
\end{eqnarray}
This formula holds for $E_{0,z}\lesssim 10^7$ $[{\rm V}\cdot{\rm m}^{-1}]$ in itself.

\section{Interaction Hamiltonian}

In this section, we derive the Hamiltonian for the interaction between the radiation field, of the specific mode, and the electric dipole moments of water molecules that solvate ions.

\subsection{Radiation-water molecules interaction}

We consider part of the classical radiation field, that is, the transverse wave in the $x$--$y$ plane, which can be written as
\begin{equation}
A_x+iA_y=A_0e^{-i\phi_0}\label{eq:AxAy}
\end{equation}
under assumption A2.
Here, $A_0$ is positive and real.
The reason why $A_z$ is not contained in Eq.(\ref{eq:AxAy}) will be explained later (see the paragraph including Eqs.(\ref{eq:Az1}) and (\ref{eq:Az2})).

The Hamiltonian for the interaction between ion-solvated water and the transverse radiation vector field $\vec{A}$ (i.e., the electromagnetic field in the radiation gauge for which the scalar potential is set to zero) can be written using the total relevant electric charge current $\vec{\widehat{j}}$ of the solvated ions and water molecules as
\begin{equation}
\widehat{H}_{{\rm int}}^{(r-i-w)}=-\sum_{I=1}^{N}\sum_{i=0}^{n_I}\vec{A}\cdot {\vec{\widehat{j}}}_{{I,i}}\;.\label{eq:Hint}
\end{equation}
Here, the natural number $n_I$ is the number of water molecules solvating the $I$-th ion.
The total relevant electric polarization current $\vec{\widehat{j}}$ is the sum of the ions' electric charge currents $\vec{\widehat{j}}_I$ and the $1$- and $2$-components of the water molecules' electric polarization currents $\dot{\vec{\widehat{d}}}_{I,i}$:
\begin{eqnarray}
\vec{\widehat{j}}_{I,0}&=&\vec{\widehat{j}}_I\;,\\
\vec{\widehat{j}}_{I,i}&=&(\vec{e}_1 \dot{\widehat{d}}_1+\vec{e}_2 \dot{\widehat{d}}_2)_{I,i}\ \ (i\neq 0)\;.\label{eq:dd}
\end{eqnarray}
Here, $\vec{\widehat{d}}_{I,i}$ is truncated and the time derivative $\dot{\vec{\widehat{d}}}_{I,i}$ is taken in the interaction picture.
In the interaction Hamiltonian (\ref{eq:Hint}), the odd-parity operator vector $\vec{\widehat{j}}_{I,i}$ ($i\neq 0$) has only off-diagonal matrix elements in the representation where the water molecule's free Hamiltonian in the two-level approximation is diagonal.

The radiation-water molecule part of $\widehat{H}_{{\rm int}}^{(r-i-w)}$ can be written as
\begin{eqnarray}
\widehat{H}_{{\rm int}}^{(r-w)}=-\sum_{I=1}^{N}\sum_{i=1}^{n_I}\vec{{A}}\cdot(\vec{e}_1 \dot{\widehat{d}}_1+\vec{e}_2 \dot{\widehat{d}}_2)_{I,i}\;.\label{eq:Hint0}
\end{eqnarray}
We approximate this equation by dropping the $I$-dependence of the number $n_I$ ($I=1,2,\ldots,N$).

Now, we incorporate ion-solvation effects into the Hamiltonian (\ref{eq:Hint0}) by taking its expectation value.
In each single-ion part of Eq.(\ref{eq:Hint0}), its expectation value is taken by using the superradiant states (\ref{eq:psig}) and (\ref{eq:psie}) of the system of water molecules that solvate the relevant ion.\footnote{Within ${\cal V}_{tr}$, the quantization axis of the state of an arbitrary water molecule is chosen as the $z$-axis because the permanent electric polarization vector of each water molecule in the states (\ref{eq:psig}), (\ref{eq:psie}), and (\ref{eq:candis}) is in the $z$-direction.}

By taking the expectation value of the Hamiltonian (\ref{eq:Hint0}), the fast internal dynamics of the resonant system with respect to each ion system is averaged out, and the rest dynamical degrees of freedom of the system of water molecules are slowly varying variables $\theta_I$.
The expectation value of the Hamiltonian (\ref{eq:Hint0}) can be written as (for its derivation, see Section 4.2)
\begin{eqnarray}
\Bigl<\widehat{H}_{{\rm int}}^{(r-w)}\Bigr>_{{\cal V}_{tr}}\approx \sum_{I=1}^{N}A_0\omega_c{\overline{\Delta n}}\widetilde{d}_0^{({\rm ave})}\frac{1}{2\sqrt{3}}\sin(\theta_I+\phi)\;.\label{eq:ave}
\end{eqnarray}
Here, we have introduced the shifted phase $\phi=\phi_0+\delta$ and the rescaled permanent polarization $\widetilde{d}_0^{({\rm ave})}=\widetilde{d}_0P_z$.
The expectation value (\ref{eq:ave}) is the ponderomotive potential of $\theta_I$ created by the radiation field of the specific mode.
In this ponderomotive potential, the value of $A$ (i.e., the values of $A_0$ and $\phi_0$) is evaluated at the spatial site of the ion cluster according to assumption A2.

\subsection{Derivation of Eq.(\ref{eq:ave})}

In this subsection, we derive Eq.(\ref{eq:ave}) according to Ref.\cite{Konishi2}.

\subsubsection{Choice of the Cartesian frame $(\vec{e}_1,\vec{e}_2,\vec{e}_3)$}

First, note that the angle of rotation of $(\vec{e}_1,\vec{e}_2)$ on the plane orthogonal to $\vec{e}_3$ can be chosen arbitrarily by adjusting the phase associated with the energy eigenstates $|g\ra$ and $|e\ra$ of each water molecule (see Eq.(\ref{eq:d0})).\cite{Dover}

Consequently, we adopt the following form of the Cartesian frame $(\vec{e}_1,\vec{e}_2,\vec{e}_3)$:
\begin{equation}
\left(\begin{array}{c}\vec{e}_1 \\ \vec{e}_2 \\ \vec{e}_3\end{array}\right)=\left(\begin{array}{ccc}1&0&0 \\ 0&\cos \xi_2 &-\sin\xi_2 \\ 0&\sin \xi_2&\cos \xi_2 \end{array}\right)\left(\begin{array}{ccc}\cos \xi_1 &-\sin \xi_1&0\\ \sin\xi_1&\cos\xi_1&0\\ 0&0&1 \end{array}\right)
\left(\begin{array}{c}\vec{e}_x \\ \vec{e}_y\\ \vec{e}_z\end{array}\right)\;.\label{eq:C1}
\end{equation}
The choice of the rotation angle $-\xi_1$ ($0\le \xi_1\le \pi$) is specified by the data $\vec{e}_3$, $\vec{e}_x$ and $\vec{e}_y$.
The choice of the rotation angle $-\xi_2$ ($0\le \xi_2<2\pi$) is specified by the data $\vec{e}_3$, $\vec{e}_x$, $\vec{e}_y$ and $\vec{e}_z$.
Here, $\vec{e}_3\cdot \vec{e}_z=\cos\xi_2$ holds.

We denote the Cartesian frame obtained by the first rotation about the $z$-axis with the rotation angle $-\xi_1$ by $(\vec{e}_{x^\prime},\vec{e}_{y^\prime},\vec{e}_z)$.
Then, the phase-changes
\begin{align}
-\phi_0&\to -\phi_0+\xi_1\;,\label{eq:C2}\\
\theta+\delta &\to \theta+\delta +\xi_1\label{eq:C3}
\end{align}
accompany the change of $(\vec{e}_x,\vec{e}_y)$ to $(\vec{e}_{x^\prime},\vec{e}_{y^\prime})$.

From Eq.(\ref{eq:C1}), the explicit forms of $\vec{e}_1$ and $\vec{e}_2$ are
\begin{align}
\vec{e}_1=&\cos \xi_1\vec{e}_x-\sin\xi_1\vec{e}_y\;,\label{eq:C7}\\
\vec{e}_2=&\cos \xi_2\sin\xi_1\vec{e}_x+\cos\xi_2\cos \xi_1\vec{e}_y-\sin\xi_2\vec{e}_z\;.\label{eq:C8}
\end{align}
For $\vec{A}\equiv A_0\cos \phi_0\vec{e}_x-A_0\sin\phi_0\vec{e}_y$, we obtain
\begin{align}
\vec{A}\cdot\vec{e}_1&=A_0\cos (\phi_0-\xi_1)\;,\label{eq:C9}\\
\vec{A}\cdot\vec{e}_2&=-A_0\cos \xi_2\sin(\phi_0-\xi_1)\;.\label{eq:C10}
\end{align}

\subsubsection{Quantum mechanical calculations}

Next, we note that the energy eigenstates $|g\ra_j$ and $|e\ra_j$ ($j=1,2,\ldots,n$) appearing in Eqs.(\ref{eq:psig}) and (\ref{eq:psie}) are defined by using the quantization axis $z$.
Then, since the truncated electric dipole moment operator
\begin{equation}
\vec{\widehat{d}}=\vec{e}_1(-\widetilde{d}_0\widehat{s}^1)+\vec{e}_2(-\widetilde{d}_0\widehat{s}^2)+\vec{e}_3\widehat{0}\label{eq:C4}
\end{equation}
can be identified with the projection $((|d^1\ra+i|d^2\ra)/2,(|d^2\ra-i|d^1\ra)/2,0)$ of a vector $(|d^1\ra,|d^2\ra,|d^3\ra)=(|d^{x^\prime}\ra,\cos \xi_2|d^{y^\prime}\ra-\sin \xi_2|d^z\ra,\sin\xi_2 |d^{y^\prime}\ra+\cos\xi_2|d^z\ra)$, we obtain
\begin{align}
\la \psi^{(g)}|(\widehat{s}^1,\widehat{s}^2)|\psi^{(g)}\ra&=\la \psi^{(g)}|(\widehat{s}^{x^\prime}, \cos \xi_2 \widehat{s}^{y^\prime})|\psi^{(g)}\ra\;,\label{eq:C5}\\
\la \psi^{(e)}|(\widehat{s}^1,\widehat{s}^2)|\psi^{(e)}\ra&=\la \psi^{(e)}|(\widehat{s}^{x^\prime}, \cos \xi_2 \widehat{s}^{y^\prime})|\psi^{(e)}\ra\;.\label{eq:C6}
\end{align}

By noting the time-derivative properties
\begin{align}
\dot{\widehat{s}^1}&=-\omega_c\widehat{s}^2\;,\label{eq:C11}\\
\dot{\widehat{s}^2}&=\omega_c\widehat{s}^1\;,\label{eq:C12}
\end{align}
and Eq.(\ref{eq:dex}), we obtain
\begin{eqnarray}
-\vec{A}\cdot\la \psi^{(g)}|\vec{e}_1\dot{\widehat{d}}_1+\vec{e}_2\dot{\widehat{d}}_2|\psi^{(g)}\ra&=&\omega_c\widetilde{d}_0\vec{A}\cdot\la \psi^{(g)}|\vec{e}_1(-{\widehat{s}}^2)+\vec{e}_2{\widehat{s}}^1|\psi^{(g)}\ra\label{eq:C13}\\
&=&A_0\omega_c\widetilde{d}_0\frac{1}{2\sqrt{3}}[\cos (\phi_0-\xi_1)(\cos\xi_2\sin (\theta+\delta +\xi_1))\nonumber\\
&&+(-\cos\xi_2\sin(\phi_0-\xi_1))(-\cos(\theta+\delta+\xi_1))]\label{eq:C14}\\
&=&A_0\omega_c\widetilde{d}_0\frac{1}{2\sqrt{3}}\cos \xi_2\sin(\theta+\phi)\;.\label{eq:C15}
\end{eqnarray}
To obtain Eq.(\ref{eq:C14}), we have used Eqs.(\ref{eq:C3}), (\ref{eq:C9}), (\ref{eq:C10}), and (\ref{eq:C5}).

In a manner completely parallel to the derivation of Eq.(\ref{eq:C15}), we obtain
\begin{equation}
-\vec{A}\cdot\la \psi^{(e)}|\vec{e}_1\dot{\widehat{d}}_1+\vec{e}_2\dot{\widehat{d}}_2|\psi^{(e)}\ra=-A_0\omega_c\widetilde{d}_0\frac{1}{2\sqrt{3}}\cos \xi_2\sin(\theta+\phi)\;.\label{eq:C16}
\end{equation}
By replacing $\cos\xi_2=\vec{e}_3\cdot\vec{e}_z$ in Eqs.(\ref{eq:C15}) and (\ref{eq:C16}) with the permanent electric polarization $P_z$ given by Eq.(\ref{eq:GPVResult}), we obtain Eq.(\ref{eq:ave}).
Here, the averaged-out fast internal dynamics of the resonant system with respect to each ion system is aggregated into the values of $w_+$, $w_-$, and $P_z$.

The reason why $A_z$ does not appear in the above calculations is that $\vec{e}_1\cdot\vec{e}_z=0$, $\vec{e}_2\cdot\vec{e}_z=-\sin\xi_2$, and
\begin{eqnarray}
\la \widetilde{0}|\sin\xi_2\cos(\theta+\delta+\xi_1)|\widetilde{0}\ra&=&\la\widetilde{1}|\sin\xi_2\cos(\theta+\delta+\xi_1)|\widetilde{1}\ra\label{eq:Az1}\\
&=&0\;.\label{eq:Az2}
\end{eqnarray}

\section{Equations of Motion}

In this section, we derive the equations of motion of the system.

\subsection{Preliminary arguments}

First, we consider the case in which the moment of inertia of the water molecules cannot vary and is its average value $I^{(w)}_{\rm ave}$ exactly.
Then, since the variables $\theta_I$ are auxiliary, it can be shown that the solutions of the equations of motion of $\theta_I$ are
\begin{eqnarray}
\theta_I+\phi=\frac{\pi}{2}+n_I\pi\;.\label{eq:phi0}
\end{eqnarray}
Here, each $n_I$ is an integer.
These solutions (\ref{eq:phi0}) are compatible with monochromatic plane-wave radiation solutions only, which satisfy the equations of motion of the radiation field (\ref{eq:CE3}) and (\ref{eq:CE4}) that will be derived later.

As can be seen from this fact, to treat the general radiation solutions of the system, we need to relax the rotational energy spectrum of the water molecules in the two-level approximation, by assuming a variable internal structure, that is, vibrational modes of the water molecules.\cite{Iw}
Specifically, by a classical mechanical procedure, we change the moment of inertia of the water molecule $I^{(w)}_{\rm ave}$ to
\begin{equation}
I^{(w)}_{\rm ave}\to I^{(w)}_{\rm ave}+\Delta I^{(w)}(t)\label{eq:Iwchange0}
\end{equation}
such that $|\Delta I^{(w)}|\ll I^{(w)}_{\rm ave}$ and $|\Delta I^{(w)}|/I^{(w)}_{\rm ave}$ is not greater than the order of $10^{-2}$.\cite{Iw}
Then, due to ${\cal E}=\hbar^2/I^{(w)}_{\rm ave}=\hbar \omega_c$, the resonance angular frequency $\omega_c$ is changed to
\begin{equation}
\omega_c\to\omega_c+\Delta \omega_c(t)\;,\ \ \Delta \omega_c(t)\approx-\omega_c\frac{\Delta I^{(w)}(t)}{I^{(w)}_{\rm ave}}\;.\label{eq:Iwchange}
\end{equation}
Here, we make the identification
\begin{equation}
\Delta \omega_c=\dot{\theta}_I\label{eq:thetaIrc}
\end{equation}
because of Eq.(\ref{eq:origin}).
Note that $|\dot{\theta}_I|\ll \omega_c$.

We can summarize this argument by saying that the frequency of the $XY$-phase $\theta_I$ of the water molecules originates in the variation of the moment of inertia from its average value.

\subsection{Equations of motion}

Now, as a result of the change (\ref{eq:Iwchange}), we can write down the equations of motion of the energy spin system as the canonical equations of $\la\sum_{{\rm WM}}\widehat{H}_{\rm free}\ra_{{\cal V}_{tr}}+\la\widehat{H}_{{\rm int}}^{(r-w)}\ra_{{\cal V}_{tr}}$ with respect to the variables $\theta_I$ and their canonical conjugates $L_I$, and the equation of motion of the radiation field of the specific mode as the Maxwell equation (note that $\la \widehat{H}_{\rm free}\ra_{{\cal V}_{tr}}=-{\cal E}/3$, and the single water molecule's ground state energy is $-{\cal E}/2$).

Specifically, with the suppression of the second time derivatives of the complex amplitude of the ansatz for the transverse radiation field\footnote{This suppression is attributed to the assumption that the characteristic time for the change of the complex amplitude of $A_{x,y}$ is much longer than the radiation wave period, which is of the order of $2\pi/\omega_c\approx 1.3\cdot 10^{-12}$ [s].}
\begin{equation}
(A_x+iA_y)(z,t)=A(t)e^{i(\omega_c t-kz)}\;,
\end{equation}
for wavenumber $k$ satisfying the dispersion relation $\omega_c=ck$\footnote{Here, $(A_x+iA_y)(z,0)=A(0)e^{-ikz}$ is the seed electromagnetic field.}, and the suppressions according to the assumptions
\begin{equation}
\dot{\theta}_I\ll \omega_c\;,\ \ \biggl|\frac{\dot{A}}{A}\biggr|\ll \omega_c\;,\label{eq:ll}
\end{equation}
the canonical equations of the phase coordinates and the angular momenta of water molecules and the equation of motion of the radiation field evaluated at the spatial site of the ion cluster are
\begin{eqnarray}
\frac{n}{6}\bigl(\omega_c+\dot{\theta}_{I}\bigr)&=&\frac{L_{I}}{I^{(w)}_{\rm ave}}\;,\label{eq:th}\\
\dot{{L}}_{I}&=&-A_0{\omega_c}{\overline{\Delta n}}\widetilde{d}_0^{({\rm ave})}\frac{1}{2\sqrt{3}}\cos(\theta_I+\phi)\;,\\
-{i\omega_c}\dot{\widetilde{{A}}}e^{i\delta}&=&-\sum_{I=1}^{N}\sum_{i=1}^n\mu c^2\widetilde{j}_{{I,i}}-\sum_{i^\prime \in{\rm bulk}}\mu c^2\widetilde{j}_{{i^\prime}}\;,\label{eq:EEq}
\end{eqnarray}
where $\widetilde{{A}}=A_0e^{-i\phi}$, $\widetilde{j}=j_x+ij_y$ is a complexification of the electric charge current density, and $\mu\approx \mu_0$ is the magnetic permeability in water.
To derive Eq.(\ref{eq:EEq}), we apply the trivial formula $\square e^{i(\omega_c t-kz)}=0$, due to the dispersion relation, and Eq.(\ref{eq:ll}) to the Maxwell equation of the radiation field.

Equation (\ref{eq:EEq}) is equivalent to the pair of equations
\begin{eqnarray}
-\omega_c\dot{A}_0\sin \phi_0-\omega_c A_0\dot{\phi}\cos \phi_0&=&-\sum_{I=1}^{N}\sum_{i=1}^n\mu c^2j_{x,{I,i}}-\sum_{i^\prime \in{\rm bulk}}\mu c^2j_{x,{i^\prime}}\;,\\
-\omega_c\dot{A}_0\cos \phi_0+\omega_c A_0\dot{\phi}\sin \phi_0&=&-\sum_{I=1}^{N}\sum_{i=1}^n\mu c^2j_{y,{I,i}}-\sum_{i^\prime \in{\rm bulk}}\mu c^2j_{y,{i^\prime}}\;.
\end{eqnarray}
By substituting $j_{x,I,i}$ and $j_{y,I,i}$ into these two equations, we obtain the equations of motion, equivalent to Eq.(\ref{eq:EEq}),
\begin{eqnarray}
\dot{A}_0&=&\frac{\mu c^2 N{\overline{\Delta n}}\widetilde{d}_0^{({\rm ave})}}{V}\frac{1}{2\sqrt{3}}\la\cos (\theta_{I}+\phi)\ra_{I}\;,\label{eq:CE3}\\
\dot{\phi}&=&-\frac{\mu c^2 N\overline{\Delta n} \widetilde{d}_0^{({\rm ave})}}{A_0V}\frac{1}{2\sqrt{3}}\la \sin (\theta_{I}+\phi)\ra_{I}\;.\label{eq:CE4}
\end{eqnarray}
Here, $V$ is the volume of the system.
Note that, for the bulk water molecules, the total contributions from $\{j_{x,i^\prime}\}$ and $\{j_{y,i^\prime}\}$ in the right-hand side of Eq.(\ref{eq:EEq}) vanish because the canonical distribution has no off-diagonal matrix elements in the energy eigenbasis.

\section{Collective Instability}
\subsection{Rescaled equations of motion}

To facilitate analysis of the equations of motion, we introduce the dimensionless variables
\begin{equation}
{\cal A}_0=A_0\biggl(\frac{\alpha}{2\beta^2}\biggr)^{1/3}\;,\ \ \tau=t\biggl(\frac{\alpha\beta}{2}\biggr)^{1/3}\;,
\end{equation}
where
\begin{eqnarray}
\alpha=\frac{\sqrt{3}\overline{\Delta n}\omega_c\widetilde{d}_0^{({\rm ave})}}{nI^{(w)}_{\rm ave}}\;,\ \ \beta=\frac{\mu c^2N\overline{\Delta n} \widetilde{d}_0^{({{\rm ave}})}}{2\sqrt{3}V}\;.
\end{eqnarray}
We denote the scaled time derivative by a prime.
Then, the set of equations of motion becomes
\begin{eqnarray}
\theta_I^{\prime\prime}&=&-2{\cal A}_0\cos (\theta_I+\phi)\;,\label{eq:EOM1}\\
{\cal A}_0^\prime&=&\la \cos (\theta_I+\phi)\ra_I\;,\label{eq:EOM2}\\
\phi^\prime&=&-\frac{1}{{\cal A}_0}\la \sin(\theta_I+\phi)\ra_I\;,\label{eq:EOM3}
\end{eqnarray}
which is exactly the same as that of the conventional FEL model as described in Ref.\cite{Nature}.

These equations of motion are non-linear and many-body coupled differential equations that cannot be solved analytically.
However, we can describe the time evolution of the system qualitatively without dropping its essence of the collective instability up to the saturation regime according to the mechanism of the conventional FEL model.

\subsection{Free electron laser-like mechanism}

\begin{figure}[htbp]
\begin{center}
\includegraphics[width=0.48\hsize,bb=0 0 260 282]{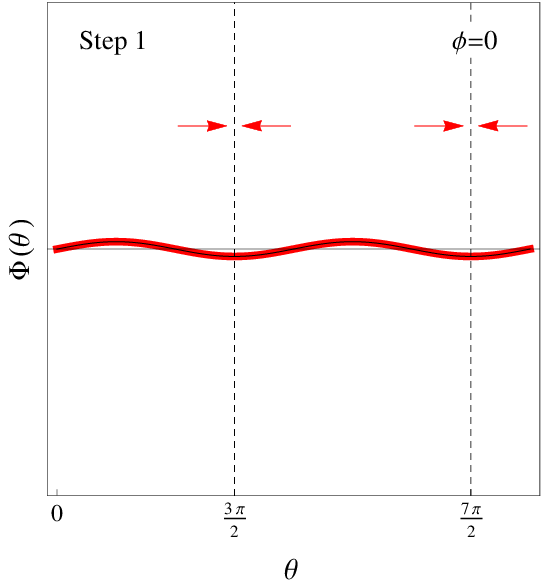}
\includegraphics[width=0.48\hsize,bb=0 0 260 282]{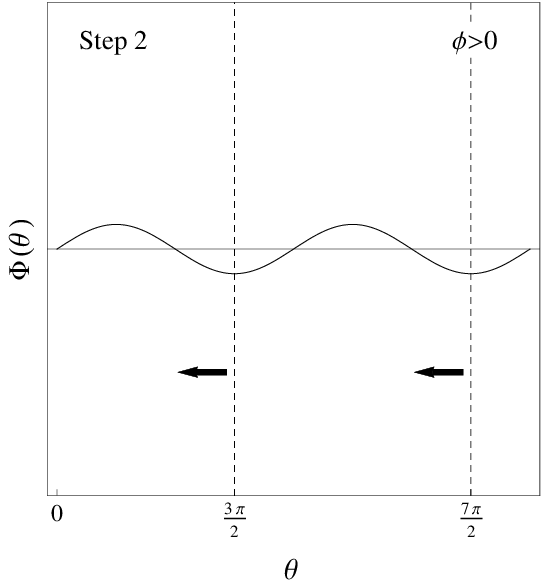}
\includegraphics[width=0.48\hsize,bb=0 0 260 282]{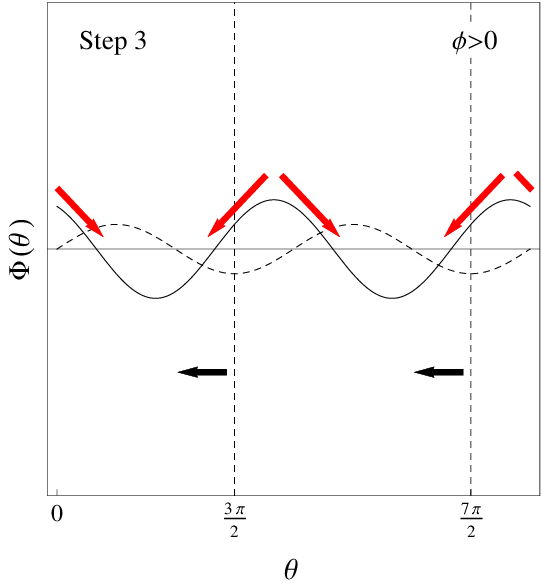}
\includegraphics[width=0.48\hsize,bb=0 0 260 282]{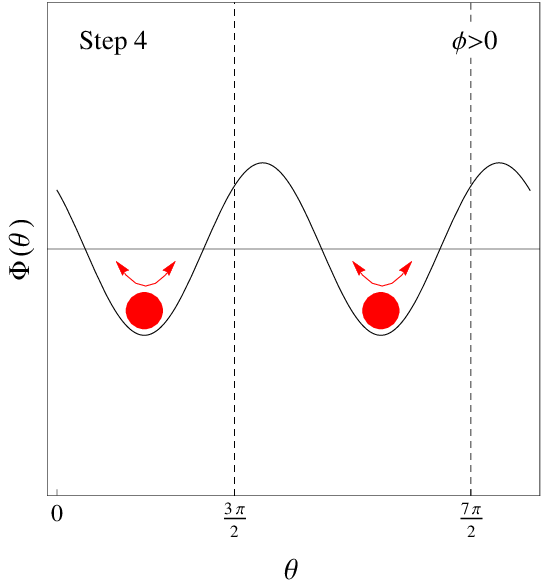}
\end{center}
\caption{Schematics of the FEL-like mechanism in four steps.
This set of figures is originally by McNeil and Thompson in Ref.18.}
\end{figure}

According to Eqs.(\ref{eq:EOM1}) to (\ref{eq:EOM3}), the FEL-like mechanism consists of the following four steps (see Fig.1).\cite{Nature}

\begin{enumerate}
\item[Step 1] Initially, the phase coordinates $\theta_I$ of the water molecules are distributed randomly and uniformly on $0\le \theta_I<2\pi$ with respect to $I$, and $\phi=0$ holds.

When the radiation intensity is initially zero, the system does not change over time due to ${\cal A}_0^\prime,\phi^\prime\approx 0$.
In this mechanism, we assume that ${\cal A}_0$ is initially small but non-zero: $0<{\cal A}_0\ll 1$.
By the force on the water molecules induced by the shallow sine curve ponderomotive potential $\Phi(\theta)=2{\cal A}_0\sin(\theta+\phi)$, this distribution becomes subtly bunched around $\theta=3\pi/2$ and $7\pi/2$.
Then, $0<-\la \sin(\theta_{I}+\phi)\ra_{I}\ll 1$ holds.

\item[Step 2] At this point, $\phi^\prime$ is positive.
Although $-\la \sin(\theta_{I}+\phi)\ra_{I}\ll 1$, since ${\cal A}_0\ll 1$, their ratio $\phi^\prime$ obtained by Eq.(\ref{eq:EOM3}) can be significant.
Then, $\phi$ grows to a positive value and the ponderomotive potential starts to move toward the left.

\item[Step 3] As a result of step 2, $\la \theta_I\ra_I+\phi\gtrsim 3\pi/2$ follows.
Then, ${\cal A}_0^\prime=\la \cos (\theta_I+\phi)\ra_I>0$ follows.
From this, the phases of the water molecules become more bunched: this is a positive feedback loop.

Intuitively, as the phases of the water molecules slide due to this accelerated movement of the ponderomotive potential and roll down to the bottom of the ponderomotive potential well, this potential well is deepened by absorbing the kinetic energies of the descending water molecules.

\item[Step 4] Finally, the potential maximally deepens (${\cal A}_0$ grows maximally as ${\cal A}_0\approx 1$), the growth of $\phi$ is quenched, and the movement of the ponderomotive potential to the left is slowed.
The maximally bunched phases of the water molecules surf the nearly static ponderomotive potential.

\end{enumerate}

\subsection{Saturated radiation intensity and gain time}

In this FEL-like mechanism, when the system reaches ${\cal A}_0\approx 1$, the positive feedback loop is expected to close; the system then enters a non-linear saturated regime.
At the same time, coherent dynamics of the radiation field and maximally bunched water molecules arise.
Namely, the quantum coherence of the water molecules is coupled over the system of ion-solvated water, and the intensity of the radiation field is magnified by a multiplicative factor of the order of $N^{4/3}$.

As the concluding formulae for this regime, we obtain
\begin{eqnarray}
\biggl(\frac{\alpha}{2\beta^2}\biggr)^{-1/3}&=&c_A\cdot \rho^{2/3}\cdot P^{1/3}_z\;,\label{eq:FA}\\
c_A&\approx&2.6\cdot 10^{-22}\ [{\rm m}^3\cdot {\rm kg}\cdot{\rm s}^{-2}\cdot {\rm A}^{-1}]\;,\\
\biggl(\frac{\alpha\beta}{2}\biggr)^{-1/3}&=&c_t\cdot \rho^{-1/3}\cdot P^{-2/3}_z\;,\label{eq:FT}\\
c_t&\approx&8.1\cdot 10^{-5}\ [{\rm m}^{-1}\cdot{\rm s}]\;.
\end{eqnarray}
Here, $\rho=N/V$ is the ion number concentration in the system.
The first formula (\ref{eq:FA}) refers to the value of $A_0$ at the time when ${\cal A}_0\approx 1$, and the second formula (\ref{eq:FT}) refers to its gain time (that is, its dynamical time).
Notably, neither formula contains the ion velocity $v$, so the ion velocity plays no essential role in our mechanism.

Finally, we have to consider saturation effects.\cite{Konishi}
Our system satisfies the condition
\begin{equation}
l_b\ll l_s=(c-v)\frac{l_g}{v}\;.
\end{equation}
Here, $l_b$ and $l_g$ are the bunch and gain lengths, respectively.
In this case, the radiation emitted by a sufficiently small bunch of water molecules could escape quickly from it due to slippage; so saturation effects would be reduced\cite{FELSR1,FELSR2}.\footnote{For the conditions for FEL superradiance, see Ref.\cite{FELSR1}.}
In a dissipative system whose original system is governed by the same equations of motion (i.e., Eqs.(\ref{eq:EOM1}) to (\ref{eq:EOM3})) as those of our system, it has been argued that superradiation, avoiding saturation effects, can be realized.\cite{FELSR1,FELSR2}
However, in our system its growth rate is too slow due to the largeness of the ratio of the slippage distance $l_s$ to the bunch length $l_b$, so this superradiance scenario for the system of all ions has to be abandoned.

\section{Illustrative Example: Action Potential Propagation in Human Brain}

In this section, we apply our mechanism for the coherence of radiation to the system for the propagation of an action potential mediated by the electric charge currents of water-solvated sodium ions Na$^+$ in typical myelinated (i.e., insulated by myelin sheathing) neuronal axons in the human brain (see Fig.2).\cite{Kandel,Hodgkin,Lodish}

\begin{figure}[htbp]
\begin{center}
\includegraphics[width=0.8\hsize,bb=0 0 260 100]{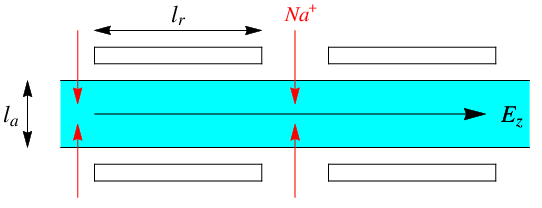}
\end{center}
\caption{Schematic of settings for action potential propagation in a myelinated neuronal axon.}
\end{figure}

The typical diameter $l_a$ of axons in the central nervous system is $10$ [$\mu$m],\cite{Axon} much shorter than the wavelength of the radiation in the resonant interaction $l_c$.\footnote{The dimension of the ion cluster in the $z$-direction does not have a definite scale.
However, to make assumption A2, we need only change the scale of $N$ and partition the ion cluster in the $z$-direction slightly.
In the following, we assume that assumption A2 is valid for this system.}
The total number of sodium ions that migrate in during an action potential event is estimated to be $Nn_{{\rm ms}}\sim 10^6$,\cite{Tegmark} where $n_{{\rm ms}}$ (estimated to be between $50$ and $100$)\cite{Kandel} is the number of myelin sheaths on one myelinated axon with an assumed length of $10$ [cm].
The conduction velocity $v$ of action potential propagation along a myelinated axon is up to $150$ [m$\cdot$s$^{-1}$].\cite{Hodgkin}
When we approximate $E_{0,z}$ to be uniform in the $z$-direction along each myelin sheath, with run length $l_r\sim 1$ [mm]\cite{Lodish} for the electric potential sloping toward the $z$-direction, it is estimated to be $\Delta U/1$ [mm] $\approx 100$ [V$\cdot$m$^{-1}$] for an electric potential difference of $\Delta U\equiv U_1-U_0\approx 0.1$ [V] between the neural firing state, with membrane electric potential $U_1\approx 0.03$ [V], and the resting state, with membrane electric potential $U_0\approx -0.07$ [V].\cite{Axon}

In this system, by setting $V \approx \pi l_a^2l_r/4$ ($\rho\approx 1.3\cdot 10^{-1}$ [$\mu$m$^{-3}$]) and $P_z\approx 4.9\cdot 10^{-7}$ according to the formulae (\ref{eq:Pzf1}) and (\ref{eq:Pzf2}), we obtain
\begin{eqnarray}
\biggl(\frac{\alpha}{2\beta^2}\biggr)^{-1/3}&\approx&5.1\cdot 10^{-13}\ [{\rm m}\cdot {\rm kg}\cdot {\rm s}^{-2}\cdot{\rm A}^{-1}]\;,\label{eq:NUM1}\\
\biggl(\frac{\alpha\beta}{2}\biggr)^{-1/3}&\approx& 2.6\cdot 10^{-6}\ [{\rm s}]\;.\label{eq:NUM2}
\end{eqnarray}
Here, the gain time (that is, the dynamical time) of our mechanism for radiation coherence is of the order of the dynamical time scale of an action potential propagation $1\ [{\rm mm}]/v\approx 6.7\cdot 10^{-6}$ [s].
From Eq.(\ref{eq:NUM2}), we also confirm that our basic approximation $|\dot{\theta}_I|\ll \omega_c$ is valid in this system.

\section{Conclusion}

In this paper, we have studied a model of a resonantly interacting system of radiation and ion cluster-solvated rotating water molecules, which have subtly variable moments of inertia under a two-level approximation.
In this model, the dimensions of the ion cluster are much shorter than the wavelength of radiation in the resonant interaction.
Furthermore, all ions in the cluster carry the same electric charge and move with very low, non-relativistic velocities in the direction parallel to an applied unidirectional static electric field.

In contrast to the bulk system of water molecules, the system of water molecules solvating an ion can be described by a pure state and a quantum $XY$ energy spin system where the expected $Z$ energy spin component is a constant.
The dynamical variable of an $XY$ energy spin is part of its phase and an {\it ion-solvated water molecule's $XY$-phase} is defined such that the angular frequency in this phase is proportional to the variation of the moment of inertia, where the proportionality constant is determined by the resonance condition (see Eqs.(\ref{eq:Iwchange}) and (\ref{eq:thetaIrc})).

Our main result is that the equations of motion of the $XY$-phases of ion-solvated water molecules, over all ions, and the transverse electromagnetic field are the same as the equations of motion of a conventional FEL.
This result leads to a mechanism for the coherence of radiation, induced by collective instability in the wave-particle interaction.

In the derivation of our result, a key earlier result is the existence of a significant permanent electric polarization $P_z$ of water molecules under a strong static electric field applied in one direction.\cite{GPV}
This is because, in the absence of this permanent electric polarization $P_z$, the FEL-like mechanism does not work: the saturated radiation intensity vanishes and the gain time (that is, the dynamical time) diverges.

As an illustrative example, we applied our mechanism to action potential propagation in myelinated neuronal axons of the human brain.
With $\rho\approx 1.3\cdot 10^{-1}$ [$\mu$m$^{-3}$] and $P_z\approx 4.9\cdot 10^{-7}$, it is found that the timescale of the mechanism is $t\approx 2.6\cdot 10^{-6}$ $[{\rm s}]$, which is of the same order as the timescale of action potential propagation.

In conclusion, our quantum coherence mechanism has desirable properties when it is applied to models of the cluster current of a large number of ions solvated in water.
As described earlier, it is assumed that a strong static electric field is applied parallel to the ion current, with the dimensions of the ion cluster being much shorter than the wavelength $l_c$ of the radiation in the resonant interaction, and the condition in Eq.(\ref{eq:ll}) must also be satisfied so that the FEL-like mechanism works.

\end{document}